\documentclass[manuscript]{aastex631}
\shortauthors{Kim et al.}

\usepackage{chemformula}
\usepackage{wasysym}
\usepackage{ulem}
\newcommand{\RomanNumeralCaps}[1]
{\MakeUppercase{\romannumeral #1}}

\received{March 8, 2022}
\revised{July 4, 2022}
\accepted{July 7, 2022}

\begin{document}

\title{Ice features of low-luminosity protostars in near-infrared spectra of \textit{AKARI}/IRC }

\author[0000-0001-8064-2801]{Jaeyeong Kim}
\affiliation{Korea Astronomy and Space Science Institute \\
            776 Daedeok-daero, Yuseong-gu \\
            Daejeon 34055, Republic of Korea}

\author[0000-0003-3119-2087]{Jeong-Eun Lee}
\affiliation{School of Space Research, Kyung Hee University \\
            1732 Deogyeong-daero, Giheung-gu \\
            Yongin, Gyeonggi-do 17104, Republic of Korea}

\author[0000-0002-2770-808X]{Woong-Seob Jeong}
\affiliation{Korea Astronomy and Space Science Institute \\
            776 Daedeok-daero, Yuseong-gu \\
            Daejeon 34055, Republic of Korea}

\author{Il-Seok Kim}
\affiliation{School of Space Research, Kyung Hee University \\
            1732 Deogyeong-daero, Giheung-gu \\
            Yongin, Gyeonggi-do 17104, Republic of Korea}

\author[0000-0003-3283-6884]{Yuri Aikawa}
\affiliation{Department of Astronomy, Graduate School of Science, The University of Tokyo \\
            7-3-1 Hongo, Bunkyo-ku \\
            Tokyo 113-0033, Japan}

\author[0000-0003-4985-8254]{Jeniffer A. Noble}
\affiliation{CNRS, Aix Marseille Universit\'e, Laboratoire PIIM \\
            UMR 7345, 13397 \\
            Marseille, France}

\author{Minho Choi}
\affiliation{Korea Astronomy and Space Science Institute \\
            776 Daedeok-daero, Yuseong-gu \\
            Daejeon 34055, Republic of Korea}
            
\author{Ho-Gyu Lee}
\affiliation{Korea Astronomy and Space Science Institute \\
            776 Daedeok-daero, Yuseong-gu \\
            Daejeon 34055, Republic of Korea}

\author[0000-0003-0749-9505]{Michael M. Dunham}
\affiliation{Department of Physics, State University of New York at Fredonia \\
            Fredonia, NY 14063, USA}

\author{Chul-Hwan Kim}
\affiliation{School of Space Research, Kyung Hee University \\
            1732 Deogyeong-daero, Giheung-gu \\
            Yongin, Gyeonggi-do 17104, Republic of Korea}

\author[0000-0002-2755-1879]{Bon-Chul Koo}
\affiliation{Department of Physics and Astronomy, Seoul National University \\
            1 Gwanak-ro, Gwanak-gu \\
            Seoul 08826, Republic of Korea}

\email{jaeyeong@kasi.re.kr} \email{jeongeun.lee@khu.ac.kr}
\correspondingauthor{Jeong-Eun Lee}


\begin{abstract}
We present near-infrared spectra of three low-luminosity protostars and one background star in the Perseus molecular cloud, acquired using the Infrared Camera (IRC) onboard the \textit{AKARI} space telescope.
For the comparison with different star-forming environments, we also present spectra of the massive protostar AFGL 7009S, where the protostellar envelope is heated significantly, and the low-mass protostar RNO 91, which is suspected to be undergoing an episodic burst.
We detected ice absorption features of \ch{H2O}, \ch{CO2}, and \ch{CO} at all spectra around the wavelengths of 3.05, 4.27, and 4.67 $\mu$m, respectively.
At least two low-luminosity protostars, we also detected the \ch{XCN} ice feature at 4.62 $\mu$m.
The presence of the crystalline \ch{H2O} ice and \ch{XCN} ice components indicates that the low-luminosity protostars experienced a hot phase via accretion bursts during the past mass accretion process.
We compared the ice abundances of low-luminosity protostars with those of the embedded low-mass protostars and the dense molecular clouds and cores, suggesting that their ice abundances reflect the strength of prior bursts and the timescale after the last burst.
\end{abstract}


\section{Introduction}
The luminosity function of young stellar objects (YSOs) with the peak luminosity lower than 1.6 $L_{\astrosun}$ raised questions about the mass accretion process \citep{eva09,dun10} since their low luminosities cannot be explained well by the standard mass accretion rate of $\sim$ 2$\times$10$^{-6}$ $M_{\astrosun}$yr$^{-1}$ \citep{shu87}.
A potential explanation of their low luminosity is the episodic accretion process, which has long-lived quiescent-accretion phases interspersed with short-lived burst-accretion phases \citep{vor05,dun10}.
In this episodic accretion model, some dynamical and chemical conditions, which cannot be explained by the current low luminosity in the continuous accretion model, should exist due to past burst accretions. Therefore, it is interesting to find observational clues supporting the episodic accretion process.

One evidence of episodic accretion can be found in the absorption features of ice spectra.
Chemistry in circumstellar envelope is a useful tracer of the history of the accretion process.
Previous studies of low- and high-mass YSOs have suggested that the thermal history of YSOs can be found from the relative distribution of molecular abundances of gas and ice components \citep{eva91,van99,lah00,boon03a,boon03b,van05,lee07,kim12,vis12}.
At the early evolutionary stage of YSOs, more molecules reside in the ice phase on dust grain surfaces rather than in the gas phase \citep{ber95,lee04}. 

If the low-luminosity objects are in the quiescent phase of an episodic accretion process, those protostars should have gone through the burst accretion phase in the past with significantly enhanced accretion rates to gain their masses. 
During the relatively short timescale of burst accretion, the abrupt heating of the surrounding dust grains should sublimate ices from the grain surfaces.
When protostars return to a quiescent phase with a low accretion rate, the dust grains cool down to the cold condition similar to that of starless or pre-stellar cores at most radii, where molecules are gradually frozen back onto the grain surfaces as ice \citep{lee07}.
In particular, significant amounts of oxygen and carbon occupy the main composition of the dust grain mantles, forming \ch{H2O}, \ch{CO}, and \ch{CO2} ices, which are the known origins of strong absorption features around near- and mid-infrared wavelengths.
The memories of hot phases must have been left in these ice features in the episodic accretion process although the current low-luminosity implies only a cold condition. 

The ice formation and evolution on the grain surfaces are linked with the thermal process of mixed ices in the mantles.
In previous \ch{CO} ice studies, it was noted that the \ch{H2O} and \ch{CO} can form mixed layers \citep{col03,pon03}.
The gas-phase \ch{CO} molecules are frozen as the \ch{CO} ice during the quiescent phase, and some of them turn into the \ch{CO2} ice \citep{kim11,kim12}.
A pure component of \ch{CO2} ice can be formed by the distillation process, in which \ch{CO} evaporates from a \ch{CO}$-$\ch{CO2} mixture, leaving \ch{CO2} ice in the stable phase due to lower thermal conditions ($\sim$ 20$-$30 K) within a protostellar envelope \citep{pon08b}.
In the episodic accretion process, if an accretion burst occurs and the temperature at some radii reaches conditions favoring sublimation of \ch{CO} ice, the \ch{CO} will evaporate from the mixed-ice grains, leaving behind the pure \ch{CO2} ice.
As a result, in a quiescent phase with a low luminosity after a burst accretion stops, the pure \ch{CO2} ice will be detectable.

Observations of the absorption features of these ice components have been conducted with space infrared telescopes that are sensitive enough to detect weak absorption features, even in the spectra of low-luminosity objects, without any atmospheric interference.
The \textit{Spitzer} IRS spectrum of \ch{CO2} ice at 15.2 $\mu$m toward low-luminosity objects shows the double-peaked feature, which is direct evidence of pure \ch{CO2} ice \citep{kim12}.
The double-peaked feature of \ch{CO2} ice indicates that the low-luminosity protostellar system must have had periods of hot phases.
The absorption features of the most abundant ices (\ch{H2O}, \ch{CO2}, and \ch{CO}) have been also observed by the Infrared Camera (IRC) aboard the \textit{AKARI} space telescope in the near-infrared range (2.5$-$5.0 $\mu$m).
This near-infrared range also covers the 4.62 $\mu$m \ch{XCN} feature \citep{gib04}, which is an important tracer of thermal history in circumstellar material.

In this paper, we study the ice abundances of the low-luminosity protostellar envelops as well as the parental cloud in the Perseus star forming region, using spectroscopic observations with the \textit{AKARI} space telescope.
For the comparison of ice abundances between the low-luminosity sources and typical embedded low- and high-mass YSOs, we also examine the near-infrared spectra of RNO 91 and AFGL 7009S, which were covered as part of our ice survey.
From the spectral analysis of the detected ice absorption features, including the \ch{XCN} ice, we explore the chemical differences in ices of embedded low-mass YSOs, molecular clouds, and low-luminosity protostars.
In Section 2, our targets, including low-luminosity sources and two other protostars, are described.
We explain the data reduction pipeline in Section 3.
In Section 4, we describe the procedure used to extract ice features from the \textit{AKARI} IRC spectra and derive their optical depth profiles and column densities.
In section 5, we discuss the \ch{XCN} ice feature detected in our low-luminosity targets as a direct evidence of past enhanced luminosity phase due to an accretion burst.
Different distributions of the ice abundances in various evolutionary stages of star formation, resulting from the surface chemistry during the different timescales after the burst ends, are also discussed.

\section{Observations}
\subsection{Targets in the Perseus molecular cloud}
The observations of three low-luminosity protostars and one background star in the Perseus region were carried out as part of the \textit{AKARI} Phase-3 (post-helium mission) open time program "ices in Star FormIng CorEs (SFICE, PI: Jeong-Eun Lee)", which observed 16 sources covering known YSOs and background stars to study physical and chemical properties at various evolutionary stages of low-mass star formation.

Our main targets for this study are three embedded protostellar objects with internal luminosity lower than 1 $L_{\astrosun}$ by \citet{dun08}.
The identification numbers of those targets as given by \citet{dun08} are 060, 090, and 104, but for this work, we label them Perseus 1, 2, and 3, respectively, because all three sources located in the Perseus molecular cloud.
The internal luminosity ($L_{int}$) of a central protostar was determined from the flux at 70 $\mu$m \citep{dun08}.
We list the source information in Table ~\ref{tab-1}, including the flux at infrared wavelengths \citep{skr06,eva03}, the luminosity derived from the 70 $\mu$m flux, and the bolometric temperature.

The environmental effects, such as the external radiation field and the cloud initial density, temperature, and chemical conditions could be ignored because all the sources are located in the same molecular cloud. 
A background star behind the Perseus molecular cloud was also included in our program to study the chemical evolution of the circumstellar material around low-luminosity protostars from their nascent condition. 
The ice compositions of a molecular cloud can be studied via the absorption features against a background infrared continuum source.
The background source used in this work is a K-type giant, and its fluxes at infrared wavelengths have been calculated from 2MASS and $Spitzer$ observations \citep{skr06,eva03}.

\subsection{Targets for comparisons: RNO 91 and AFGL 7009S}
RNO 91 and AFGL 7009S are normal low- and high-mass protostars, respectively, and their spectroscopic studies have been carried out in the mid-infrared ranges.
We use these two protostars for comparisons with our low-luminosity targets in the view of chemical evolution.

RNO 91 (also known as IRAS 16316-1540) is known as a Class I protostar \citep{che09} embedded in a reflection nebula \citep{may07}, located in the Ophiuchus molecular cloud.
It was recently observed as part of the protoplanetary disk survey with the Immersion Grating Infrared Spectrograph (IGRINS) at the near-infrared $H$ and $K$ bands \citep{yoo21}.
The IGRINS spectra of RNO 91 show broad absorption features in the \ch{CO} overtone bands, water vapor bands, and various atomic lines, which are similar to the spectral characteristics of FU Orionis objects \citep{lee15a,park20}. In addition, it shows the double-peaked pure CO ice feature at 15 $\mu$m \citep{pon08b}.
Therefore, those features strongly suggest that RNO 91 is undergoing an accretion burst with an enhanced mass accretion rate in spite of its normal luminosity.
On the other hand, the low luminosity sources are considered in their quiescent  phases with low accretion rates \citep{dun10}. 
In the episodic accretion model, all protostars even with current low luminosities must have undergone burst accretions. 
In addition, the chemical compositions affected by the burst accretions remain much longer than the dynamical timescale of burst accretion \citep{lee07}. 
Therefore, the comparisons of the ice compositions between RNO 91 and the low-luminosity sources will provide an important opportunity to investigate the chemical evolution in the episodic accretion model.

AFGL 7009S (also known as IRAS 18316-0602) is one of the most massive class \RomanNumeralCaps{1} YSOs studied to date \citep{gib04}.
It has been observed with the Infrared Space Observatory (ISO) Short and Long Wavelengths Spectrometers (SWS and LWS), and numerous studies for this target have been carried out \citep{dhen96,dar98,dar00,gib04}.
This object has saturated absorption features of \ch{H2O} and \ch{CO2} ices throughout 3 and 4.27 $\mu$m, as observed from the ISO.
The \ch{H2O} column density was estimated by \citet{dhen96} using both the 6 and 13 $\mu$m features.
Our near-infrared spectroscopic data for AFGL 7009S can provide a comparison of ice chemistry between low- and high-mass star formation in the deeply embedded stage.

\subsection{Observations with \textit{AKARI} Infrared Camera (IRC)}
\begin{figure}
\begin{center}
\includegraphics[scale=0.7]{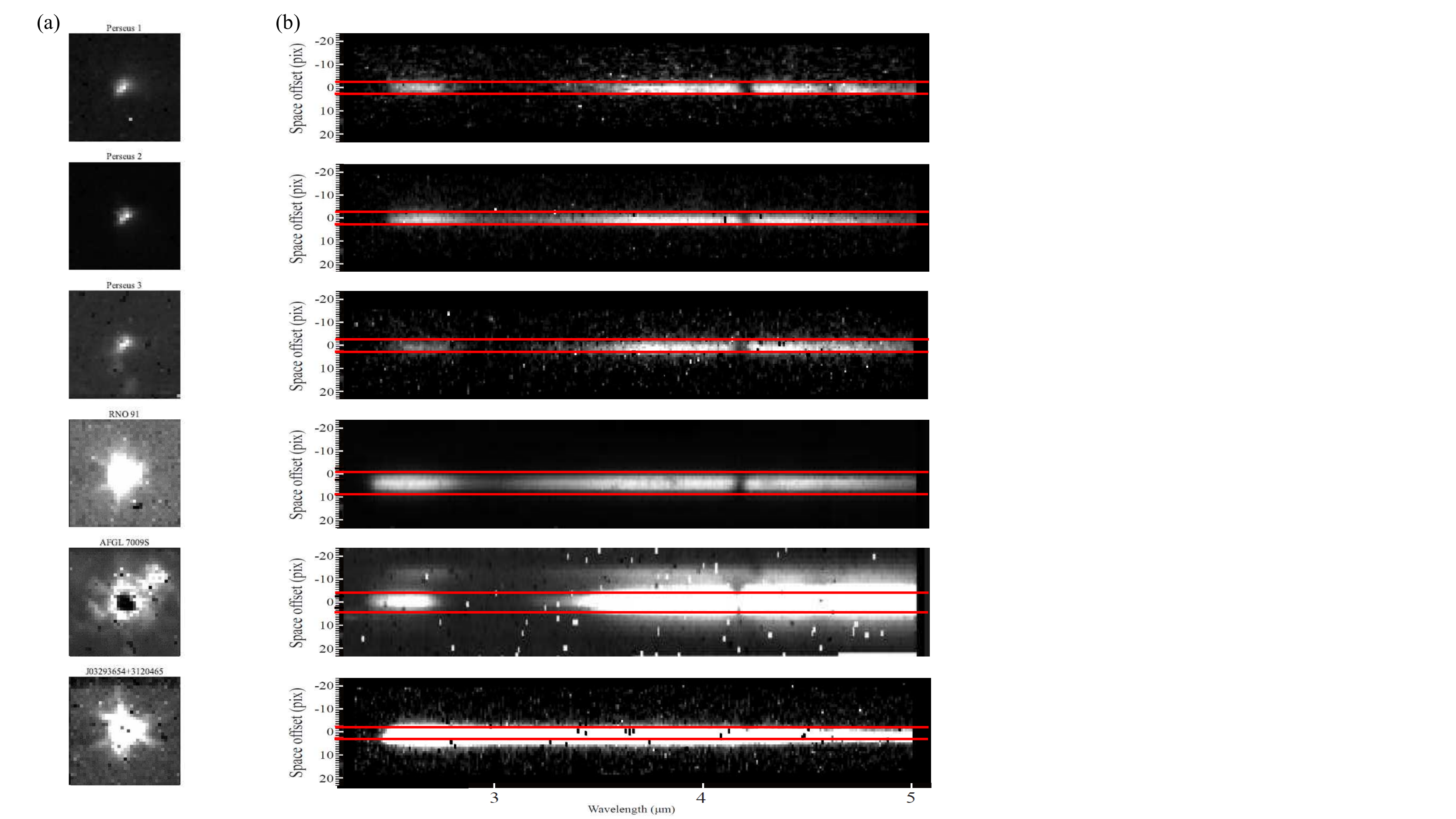}
\caption{a) Images of our targets with the N3 filter of 1$\arcmin$ $\times$ 1$\arcmin$ field of view. The coordinate at the center of each image is listed in Table~\ref{tab-1}. b) Spectral images of our targets, obtained from the spectroscopic pipeline process. Dispersion direction is left to right. Red lines indicate the pixel range of spatial integration.}\label{fig1}
\end{center}
\end{figure}

We performed near-infrared spectroscopic observations of our targets using the \textit{AKARI} IRC on February 15 and 17, August 21 and 23, September 2, and October 1 in 2009.
We used the IRC grism (NG) mode, which covers 2.5$-$5.0 $\mu$m with a spectral resolution of R $\sim$ 100 at 3.6 $\mu$m \citep{ona07}.
This resolving power was taken from the dispersion rate per pixel (1.46$\arcsec$) of 0.0097 $\mu$m \citep{ohy07} at the focal plane, and such resolution range has been guaranteed to detect the ice absorption bands \citep{aik12,nob13}.
We adopted the standard point source spectroscopic (Np) mode, which provides a 1$\arcmin$ $\times$ 1$\arcmin$ slit aperture to minimize light contamination by nearby bright sources.
From the standard spectroscopic observing sequence, we obtained seven spectroscopic frames and one imaging frame (reference image) in the middle of them.
Each frame consists of a short-exposure (4-second) and a long-exposure (44-second).
Combining the spectroscopic frames, the total integration time provided 400-second exposures.
Figure ~\ref{fig1}(a) shows the imaging frames of our targets.

\section{Data Reduction}\label{sec:Reduction}
A data reduction process was performed through the spectroscopy pipeline for $\textit{AKARI}$ IRC phase 3 \citep{ohy07,ona07}.
This pipeline was provided by the $\textit{AKARI}$ project team, and we used its latest version of the toolkit, IRC\_SPECRED, based on the IDL environment.
For dark subtraction from the raw data, five self-dark images were taken at the preceding and subsequent frames of the data archive.
We applied a linearity correction and saturation masking by comparing the reference images between short and long exposure frames.
A flat fielding process was not included during the data reduction due to poor signal-to-noise ratios of a flat image of NG spectra in the Np mode.
Subtraction of the sky values was also performed using the dark-subtracted image.

The source detection of the reference image was executed through 'DAOFIND' task automatically or manually, by inserting source coordinates into the pipeline command.
Based on the source coordinate found in the reference image before stacking into a final spectral image, the eight spectral images were shifted by the attitude drift of the $\textit{AKARI}$ satellite.
After subtracting the cosmic rays and the remaining background signals, the spectral images were stacked and averaged with a wavelength calibration for the reference image and the extracted spectrum.
The signal was integrated over five pixels along with the spatial direction.
Figure ~\ref{fig1}(b) shows the spectral images at each target.
The stellar light was dispersed from left (short wavelengths) to right (long wavelengths).
The spectrum was not perfectly aligned with the rows of pixels, and we adopted the correction formula to tilt the spectrum as indicated below,
\begin{equation}
 Y = -0.00746929(X-X_{0})+Y_{0},\\\end{equation}
where $X$ and $Y$ are the pixel number in the dispersion direction and spatial direction, respectively, and $X_{0}$ and $Y_{0}$ indicate the center of the spectrum.

Finally, the integrated signal was corrected using a spectral response function before being converted into a spectrum.
The response function in the previous version of the toolkit was based on calibrated observations of blue standard stars, which suffered from severe 2nd-order light contamination \citep{bab16}.
In the current version of the toolkit, the response function corrects the effect of the contamination on the part considered.
The flux error was estimated using the noise level in the sky region and the response function.
According to \citet{aik12}, the linear correlation between pixel and wavelength is less certain at longer wavelengths.
Therefore, We shifted the spectral data by 1$-$2 pixels using the task CHANGE\_WAVE\_OFFSET, to fit the pixel position and spectral wavelengths.

\section{Results}
\subsection{Reduced Spectra}\label{sec:Spectra}
\begin{figure}
\begin{center}
\includegraphics[scale=0.5]{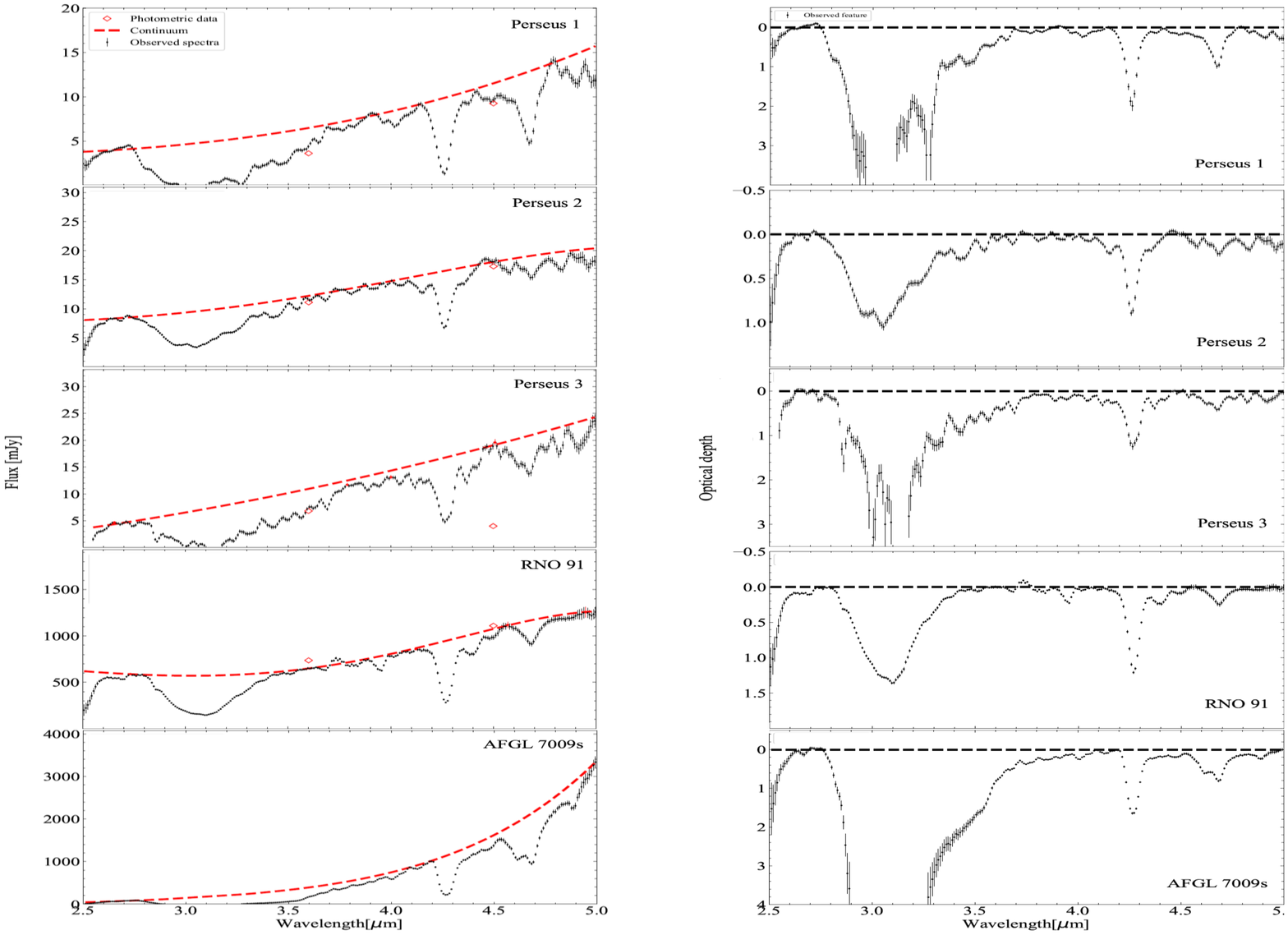}
\caption{The observed $AKARI$ IRC spectra (left) and the corresponding optical depth plot (right) for all the sources in this study, except the background star J03293654+3129465 (See Figure ~\ref{fig3}). The red dashed line shows the continuum determined by the second- to third-order polynomial fitting of the spectrum. The opened red diamonds are the \textit{Spitzer} IRAC 3.6 and 4.5 $\mu$m fluxes, which are referred from c2d YSO catalog \citep{eva03}. The optical depth is calculated from the equation, $\tau$ $=$ ln($I_{0}$/$I$), where $I_{0}$ and $I$ are the continuum and observed fluxes, respectively.}\label{fig2}
\end{center}
\end{figure}

In Figure ~\ref{fig2}, the reduced $\textit{AKARI}$ IRC spectra of all protostars are presented; the absorption features of \ch{H2O}, \ch{CO2}, and \ch{CO} ices are clearly detected.
All of our targets show deep and broad absorption features of \ch{H2O} ice in the wavelength range of 2.7 to 3.4 $\mu$m.
In the case of AFGL 7009S, strong extinction towards the source saturates the absorption feature throughout the wavelengths from 2.7 to 3.6 $\mu$m, including \ch{H2O} ice.
Other ice features, such as \ch{CH4} \citep{lac91,boo04} and \ch{CH3OH} \citep{gri91,bro96}, were observed at 3.3$-$3.5 $\mu$m, but it is difficult to extract their absorption profile from the blended features due to the low spectral resolution of the $\textit{AKARI}$ IRC.
The absorption feature of \ch{CO2} ice around 4.27 $\mu$m was clearly detected toward all targets.
At the wavelength around 4.6 $\mu$m for Perseus 1 and 3, RNO 91, and AFGL 7009S, there is a hint for another ice component overlapping with the \ch{CO} absorption feature at 4.67 $\mu$m.
Many near-infrared observations have revealed the same feature on the blue wing part of the \ch{CO} absorption feature \citep{teg95,chi98,whi01,van05,aik12}, which was suggested as the absorption feature of the \ch{XCN} ice.
\citet{lac84} and \citet{pen99} reported that the 4.62 $\mu$m absorption feature of \ch{XCN} ice consists of a nitrile group and an unknown component $'$X$'$. 
Many laboratory studies suggested that UV photolysis or cosmic ray irradiation of ice mantle could make solid state \ch{OCN-} on grain surfaces \citep{lac84,gri87,ber00,pal00,hud01,van04}.
In addition to these ice components, there are some absorption features around 4.8 and 4.9 $\mu$m.
For Perseus 3 and the background star, the absorption features having a peak position around 4.78 $\mu$m are likely associated with the $^{13}$CO ice \citep{boo02,pon03}.
We also detected another absorption feature at 4.83 $\mu$m toward the low-luminosity targets.
However, we could not find any corresponding ice features from previous studies.
The 4.9 $\mu$m absorption feature detected toward all targets was identified as the \ch{OCS} ice.
The \ch{OCS} ice can be produced when the interstellar ices containing \ch{CO} and \ch{CO2} are exposed to UV photons or cosmic rays \citep{pal97}.

\subsection{Analyses of the ice absorption features}\label{sec:IceFeature}
In order to quantify the ice abundances from the absorption features, we needed to derive the column density ($N$) of each ice component, which were obtained using the following equation,
\begin{equation}
 N = \frac{\int \tau d\nu}{A}\\\end{equation}
where $A$ is the band strength (in cm molecules$^{-1}$) and $\tau$ is the optical depth.
The band strengths of \ch{H2O}, \ch{CO2}, and \ch{CO} are 2.0$\times$10$^{-16}$, 7.6$\times$10$^{-17}$, and 1.1$\times$10$^{-17}$ cm molecule$^{-1}$, respectively \citep{ger95}.
We determined the optical depth of the spectrum through the normalized flux using the continuum, which is interpolated with a polynomial function \citep{boo08,pon08b}.
It was not easy to select continuum points to be used as pristine continuum levels because there are numerous absorption and emission bands in our spectral window.
We used the second- to third-order polynomial functions to fit the continuum using five wavelength ranges: 2.6$-$2.8, 3.4$-$3.7, 4.1$-$4.2, 4.4$-$4.6, and 4.7$-$4.9 $\mu$m.
Those fitting ranges were set to avoid absorption features, varying slightly with objects to optimize continuum levels of spectrum.
The optimized continuum at each target are overlaid as the dashed curve in Figure ~\ref{fig2}.

\begin{figure}
\begin{center}
\includegraphics[scale=0.5]{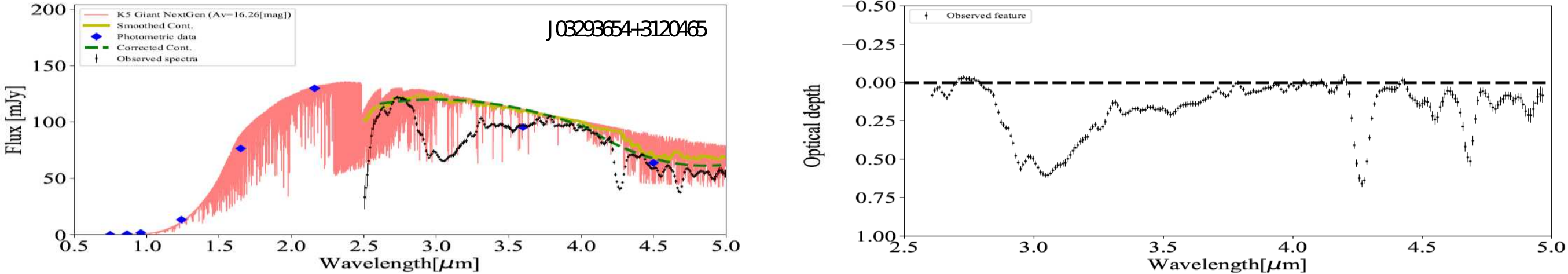}
\caption{The same plots as Figure ~\ref{fig2}, but for the background star behind the Perseus molecular cloud. The green dashed line shows the corrected continuum determined by the reddened (red line) and smoothed NextGen model data (gold line) and 2MASS data points (blue squares).}\label{fig3}
\end{center}
\end{figure}

In the case of the background star, a different method was employed to determine the optical depth of its spectral data.
We adopted a NextGen stellar model as the continuum of the background star, a K5 giant, to remove the effect of the absorption feature in the intrinsic stellar spectrum.
Foreground dust extinction ($A_\mathrm{V}$) of the background star was estimated using a color-color diagram from the 2MASS photometric data \citep{cut03}.
According to the opacity and reddening vector of 2.078 \citep{wei01}, we derived $A_\mathrm{V}$ (16.25 mag) and applied it to the model data.
The reddened model data were fitted to our spectrum and then were smoothed to the spectral resolution of \textit{AKARI} IRC.
For the case of wavelengths shorter than 3.7 $\mu$m, we extrapolated the smoothed model data with the 2MASS data points, due to the contamination by the broad absorption feature around 3.0 $\mu$m.
At longer wavelengths, we used a third-order polynomial function to fit the continuum levels.
Finally, the continuum data were produced by interpolating them.
Figure ~\ref{fig3} shows the corrected continuum curve and smoothed NextGen model data on top of the spectrum of the background star.

Derivation of the intrinsic optical depth profile for each ice component is conducted via the fitting of experimental ice absorbance data to the observed spectra.
We used the experimental data of \citet{ehr96}, \citet{ger95}, and \citet{fra04} to fit our spectra and calculate the ice column densities.
However, in the near-infrared range (2.5$-$5.0 $\mu$m), the grain shape and size can affect the shape and peak position of the ice absorption features \citep{ehr97,nob13}.
For the experimental data of the \ch{CO2} and \ch{CO} ice components, we adopted the continuously distributed ellipsoids (CDE) grain model to correct their absorbance profiles because the absorption wavelengths are similar to the sizes of the grains \citep{ehr97,pon03}.
After the grain shape correction, given the low spectral resolution of IRC, we convolved the experimental data with the $\textit{AKARI}$ IRC instrumental line profile resulted from the point spread function (PSF) of the detector, and then re-binned it to the $\textit{AKARI}$'s resolving power.
The experimental data are divided into polar and apolar ices, and thus, we used combinations of polar and apolar mixtures to fit those observed absorption features by weighting the ice abundances uniformly at all wavelengths.
The best-fit experimental profiles are listed in Table ~\ref{tab-2}.

\begin{figure}
\begin{center}
\includegraphics[scale=0.61]{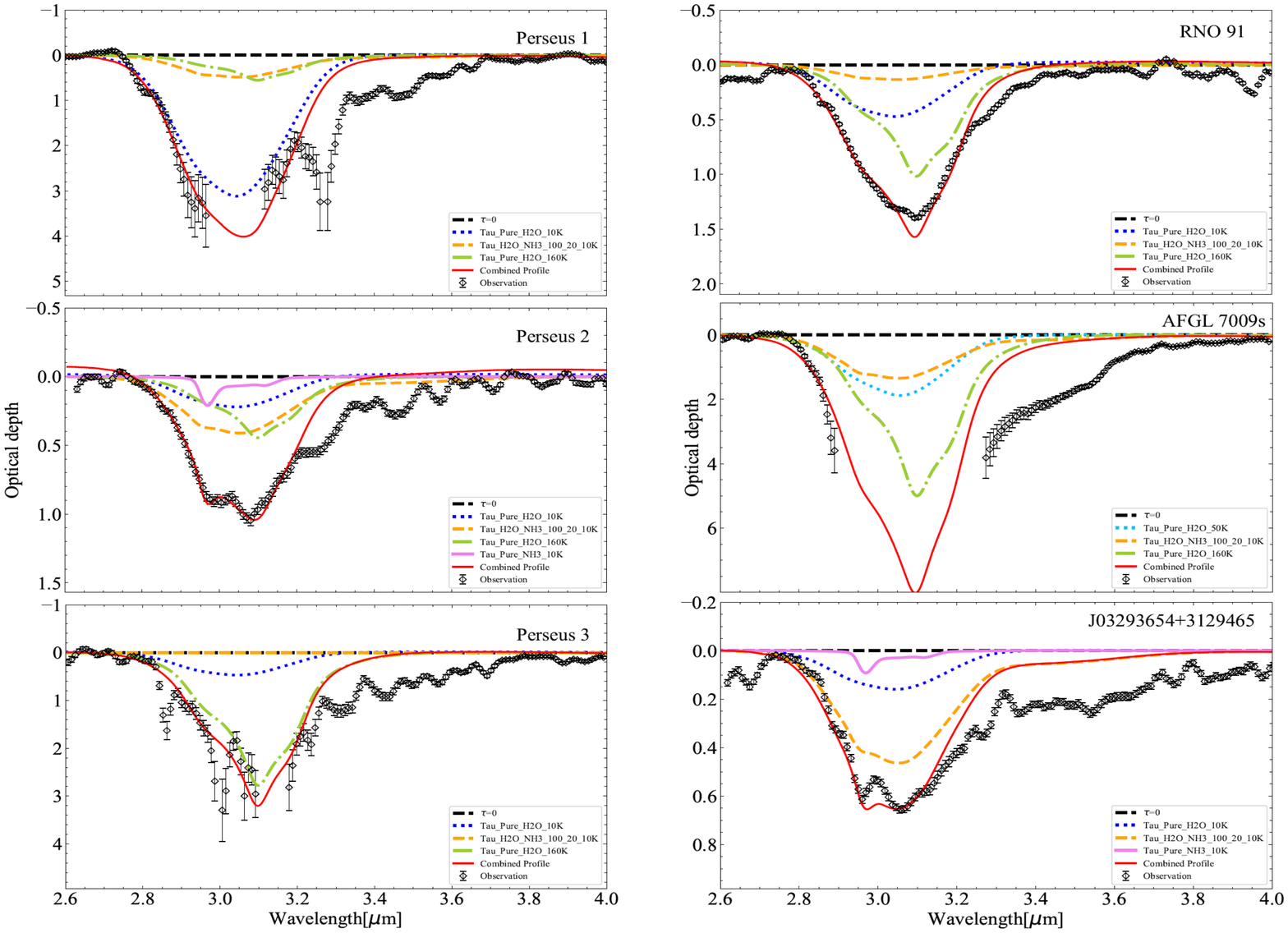}
\caption{The best-fit results of the \ch{H2O} ice absorption feature for all the sources in this study. Black diamonds present the derived data points to the optical depth with error bars. Some water ice features are saturated (Perseus 1 and 3, and AFGL 7009S). The fitted laboratory profiles are described in the bottom right of each panel. Combined ice profile with the pure amorphous (blue or cyan dotted lines) and crystalline (green dot-dashed line) components and polar mixture (orange dashed line) is described as the red solid line. Pure \ch{NH3} ice profile (magenta) is applied to fit double-peaked features at Perseus 2 and the background star.}\label{fig4}
\end{center}
\end{figure}

Figure ~\ref{fig4} shows the best-fit results for the absorption features of \ch{H2O} ice.
We used the wavelength range of 2.78$-$2.95 $\mu$m to evaluate the goodness of fitting of the \ch{H2O} absorption features, first.
We avoided the central data points of the \ch{H2O} ice absorption from the experimental profile because half of our targets are affected significantly by saturation (see Perseus 1, Perseus 3, and AFGL 7009S).
The absorption features of the protostars were fitted by a mixture of amorphous and crystalline pure \ch{H2O} ices at 10 K and 160 K, respectively.
The crystalline ice component, which has a sharp absorption feature at 3.1 $\mu$m, is linked to thermal processing of the ices \citep{boo15}.
In the case of low-luminosity protostars, the existence of the crystalline ice component may support that they have experienced higher luminosities inducing crystallization of amorphous \ch{H2O} ices.
Interestingly, the \ch{H2O} absorption feature toward Perseus 3 was fitted mostly by the crystalline ice component, while Perseus 1 was fitted mostly by the amorphous component.

Together with the water ice, we also considered a polar ice mixture of \ch{H2O}:\ch{NH3} $=$ 100:20 at 10 K to fit the absorption by the \ch{NH3} ice at 2.98 $\mu$m detected in Perseus 2 and the background star. 
The \ch{NH3} absorption features were previously detected at 2.98 \citep[pure;][]{aik12} or 3.47 \citep[hydrate;][]{dar02} $\mu$m.
Although the absorption spectra of Perseus 2 and the background star show a double-peaked profile at the central part, the peak positions, therefore, their origins are different; for Perseus 2, the peak is located at 3.10 $\mu$m, and thus, produced by the crystalline \ch{H2O} ice component, while for the background star, the absorption peak appears at 3.05 $\mu$m, indicative of the amorphous \ch{H2O} ice component.
Therefore, we fitted the ice feature of the background star using a combination of the amorphous and the \ch{NH3}-mixed profiles at 10 K.
For AFGL 7009S, we applied a higher temperature of 50 K to the amorphous ice profile, according to the previous study of the same target \citep{gib04}.
We fitted the absorption feature of AFGL 7009S by adopting the results of \citet{gib04} because it was challenging to evaluate the goodness of fitting due to the extremely saturated \ch{H2O} absorption.

\begin{figure}
\begin{center}
\includegraphics[scale=0.25]{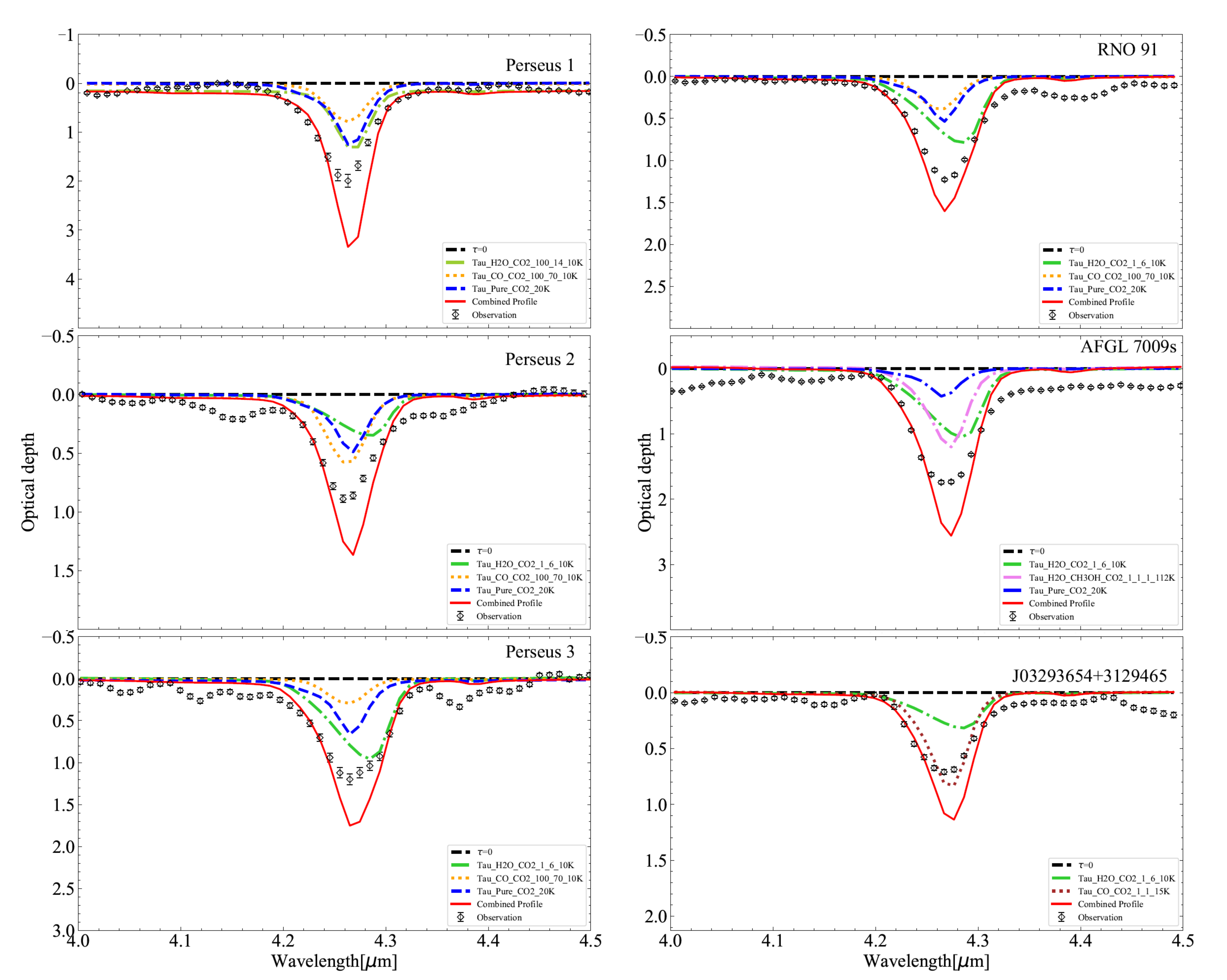}
\caption{The best-fit results of the \ch{CO2} ice absorption feature for all the sources in this study. Pure \ch{CO2} ice profile at 20 K is applied to fit each ice absorption features, except the background star. The fitted laboratory profiles are described in the bottom right of each panel. Combined ice profile with the pure (blue dashed line), \ch{H2O}-mixed (green dot-dashed line), and \ch{CO}-mixed (orange dotted line) components is described as the red solid line. We fit the absorption feature of AFGL 7009S to the laboratory ice profile for a massive protostar \citep{gib04}. Brown dotted line for the background star denotes an apolar mixture of \ch{CO}:\ch{CO2} $=$ 1:1 at 15 K.}\label{fig5}
\end{center}
\end{figure}

The stretching mode of \ch{CO2} at 4.27 $\mu$m shows a clear and deep absorption feature toward our targets.
The deep absorption feature can be easily saturated \citep{nob13}, and thus, we avoided the central four data points and employed both side wings of the \ch{CO2} absorption feature to fit the experimental profiles; the fitted profiles result in a little deeper central feature compared to the observed one.
We first adopted the same ice components as applied to the 15 $\mu$m bending mode of the \ch{CO2} ice \citep{pon08b,kim12}, but it was difficult to decompose each ice component from the observed ice feature of the \ch{CO2} stretching mode probably due to the spectral resolution of \textit{AKARI}.
As a result, we used only three components from the profiles; \ch{H2O}-rich and \ch{CO}-rich ice mixtures and pure \ch{CO2} ice.
We fitted the \ch{CO2} absorption feature of Perseus 1 with the \ch{H2O}-rich (\ch{H2O}:\ch{CO2} $=$ 100:14) and \ch{CO}-rich (\ch{CO}:\ch{CO2} $=$ 100:70) ice mixtures, adopting the column density ratio of those ice mixtures derived from \citet{kim12}.
The apolar mixture of \ch{H2O}:\ch{CO2} $=$ 1:6 at 10 K was applied to fit the \ch{CO2} absorption features of other targets.
We set a different apolar mixture to fit the \ch{CO2} absorption feature of the background star, following \citet{nob13}.
For AFGL 7009S, we adopted the experimental profile of \ch{H2O}:\ch{CH3OH}:\ch{CO2} $=$ 1:1:1 at 112 K, which was used to fit the \ch{CO2} absorption feature for massive protostars \citep{gib04}.
In the quiescent phase after a burst accretion stops, some \ch{CO2} ice is left as the pure form after \ch{CO} evaporates from the \ch{CO2}$-$\ch{CO} ice mixture, as presented in \citet{kim12} using the \textit{Spitzer}/IRS spectra.
For our low-luminosity targets, we used the pure component of the \ch{CO2} ice at 20 K \citep{kim12}.
Figure ~\ref{fig5} shows the best-fit results for the \ch{CO2} ice absorption features.

\begin{figure}
\begin{center}
\includegraphics[scale=0.25]{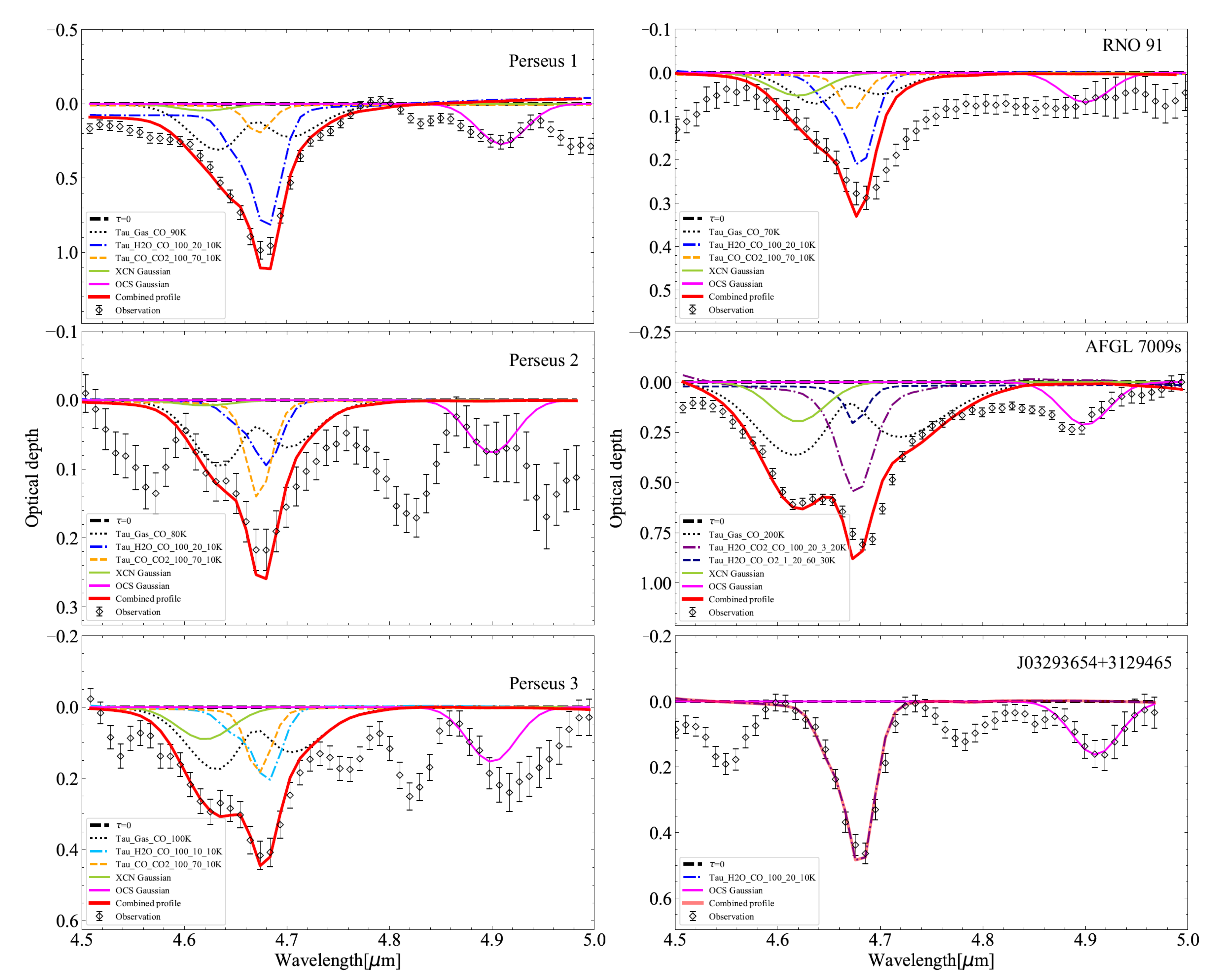}
\caption{The best-fit results of the \ch{CO}, \ch{XCN}, \ch{OCS} ice absorption features for all the sources in this study.
The polar ice (dot-dashed line) and apolar ice mixtures (orange dashed line) at 10 K compose the \ch{CO} ice features, except for AFGL 7009S.
We fit the absorption feature of AFGL 7009S to the laboratory ice profile for a massive protostar \citep{gib04}.
The green and magenta solid line show the Gaussian profiles for the \ch{XCN} and \ch{OCS} ice features, respectively.
The black dotted line shows the \ch{CO} gas profile at each relevant temperature.
The fitted laboratory profiles are described in the bottom left of each panel.
Combined ice profile is described as the red solid line.}\label{fig6}
\end{center}
\end{figure}

The absorption features of \ch{CO} ice toward the protostars generally show a broad wing structure, which is likely overlapped with other ice and gas components.
The blue wing is probably combined with the \ch{XCN} ice and absorption lines of gas phase \ch{CO} (\textit{R} branch).
The red wing also corresponds to the \textit{P} branch of the \ch{CO} gas absorption lines broaden with the low spectral resolution of \textit{AKARI} IRC.
The gas absorption feature of \ch{CO} at the warm temperature \citep[$>$70 K;][]{mit90} has been reported in embedded protostars \citep{whi97,aik12,ona21}.
Therefore, detection of the warm \ch{CO} gas component, along with the crystalline component of the \ch{H2O} ice, is considered a strong signature of heating by the central protostar \citep{ona21}.
We fitted the \ch{XCN} ice with a Gaussian profile, but the \ch{CO} gas absorption feature was fitted with the gas model profile.
The gas temperatures of the fitted \ch{CO} gas are in the range of 80 to 100 K for the low-luminosity sources, suggesting that their envelopes had a high thermal condition in the recent past.
In the case of AFGL 7009S, we applied the absorption profile corresponding to the \ch{CO} gas temperature of 200 K along with the deep Gaussian component of the \ch{XCN} ice.
After the subtraction of those components, we fitted the \ch{CO} ice absorption features of the protostars with the experimental profiles of \ch{CO}-rich (\ch{CO}:\ch{CO2} $=$ 100:70 at 10 K) and \ch{H2O}-rich (\ch{H2O}:\ch{CO} $=$ 100:20 at 10 K) ice mixtures.
Only for the case of Perseus 3, the \ch{H2O}-rich ice mixture of \ch{H2O}:\ch{CO} $=$ 100:10 at 10 K fits better than the \ch{H2O}-rich ice mixture of \ch{H2O}:\ch{CO} $=$ 100:20 at 10 K.
We used a higher temperature condition of the \ch{CO} ice \citep{gib04} to fit the absorption feature of AFGL 7009S.
In the case of the background star, only the \ch{H2O}-rich ice mixture was employed to fit its absorption feature.
We show the best-fit results for the \ch{XCN} ice and both the gas and ice absorption features of \ch{CO} in Figure ~\ref{fig6}.

Consequently, we derived the column density of each ice component for our targets using those best-fit results.
The column density errors for \ch{H2O}, \ch{CO2}, and \ch{OCS} were estimated using the optical depth errors in their absorption area at 2.80$-$3.20 $\mu$m, 4.15$-$4.35 $\mu$m, and 4.85$-$4.95 $\mu$m, respectively.
If the absorption area for the \ch{H2O} ice was saturated, such as AFGL 7009S, we adopted the error from the wavelength range between 2.8$-$2.9 $\mu$m.
We also calculated the column density error for the combined absorption of XCN and CO, dividing them into the 4.58$-$4.68 $\mu$m and 4.60$-$4.75 $\mu$m regions.

For the \ch{CO2} ice, we compared the results of Perseus 1, 2, RNO 91, and AFGL 7009S with those previously reported by the \textit{Spitzer} and \textit{ISO} spectra \citep{gib04,pon08b,kim12}. 
The \ch{CO2} ice column densities calculated from the stretching mode in this work are almost twice lower than those calculated from the bending mode except for Perseus 2. 
These differences could be attributed to the underestimated saturation effect because of the lower spectral resolution.
The relative ice abundances of \ch{CO} and \ch{CO2} were calculated with respect to the \ch{H2O} ice to avoid the effect of physical conditions such as source size.
The derived column densities and their relative ratios for our targets are summarized in Table ~\ref{tab-3} and ~\ref{tab-4}, respectively.
In our analysis, due to the saturation effect, the \ch{H2O} ice column density of RNO 91 was also underestimated compared to the prior work \citep{pon08b}.
As a result, the derived \ch{CO2} ice abundance relative to the \ch{H2O} ice is similar as $\sim$0.3 in both analyses (Table ~\ref{tab-4}).

\section{Discussion}
\subsection{The XCN ice feature}\label{sec:XCN_ice}
Detection of the double-peaked absorption feature of pure \ch{CO2} ice at 15.2 $\mu$m has been considered as a strong evidence of episodic accretion because the pure \ch{CO2} ice can be produced by the evaporation of the \ch{CO} ice from the \ch{CO}$-$\ch{CO2} ice mixture or the segregation from the \ch{H2O}$-$\ch{CO2} ice mixture on grain surfaces heated above the temperatures derived from the current luminosities \citep{lee07,pon08b,kim12}.

The \ch{XCN} ice feature detected by \textit{AKARI} IRC could be used as another evidence of episodic accretion.
In high luminosity protostars such as AFGL 7009S, the strong \ch{XCN} absorption feature has been considered as a result of the strong UV radiation from the central protostar.
Therefore, the \ch{XCN} absorption feature is not expected in low-luminosity sources.
Consequently, the detection of the absorption features at 4.62 $\mu$m (green solid lines in Figure ~\ref{fig6}) in our low-luminosity targets suggests that those low luminosity sources had energetic burst events in the past.
\citet{kim12} also detected the double-peaked feature of pure \ch{CO2} ice in Perseus 1 and 2.
RNO 91, which is believed to undergo an outburst event based on the broadened \ch{CO} overtone absorption spectra in \textit{K}-band \citep{yoo21}, shows the \ch{XCN} ice feature as well as the pure \ch{CO2} ice feature \citep{pon08b}.
Moreover, the detection of crystalline components of \ch{H2O} ice at all our low-luminosity targets also supports prior burst accretion events because the crystallization of \ch{H2O} ice reflects the thermal processing at high temperatures \citep{boo15}.

\subsection{Ice abundances}
\begin{figure}
\begin{center}
\includegraphics[scale=0.7]{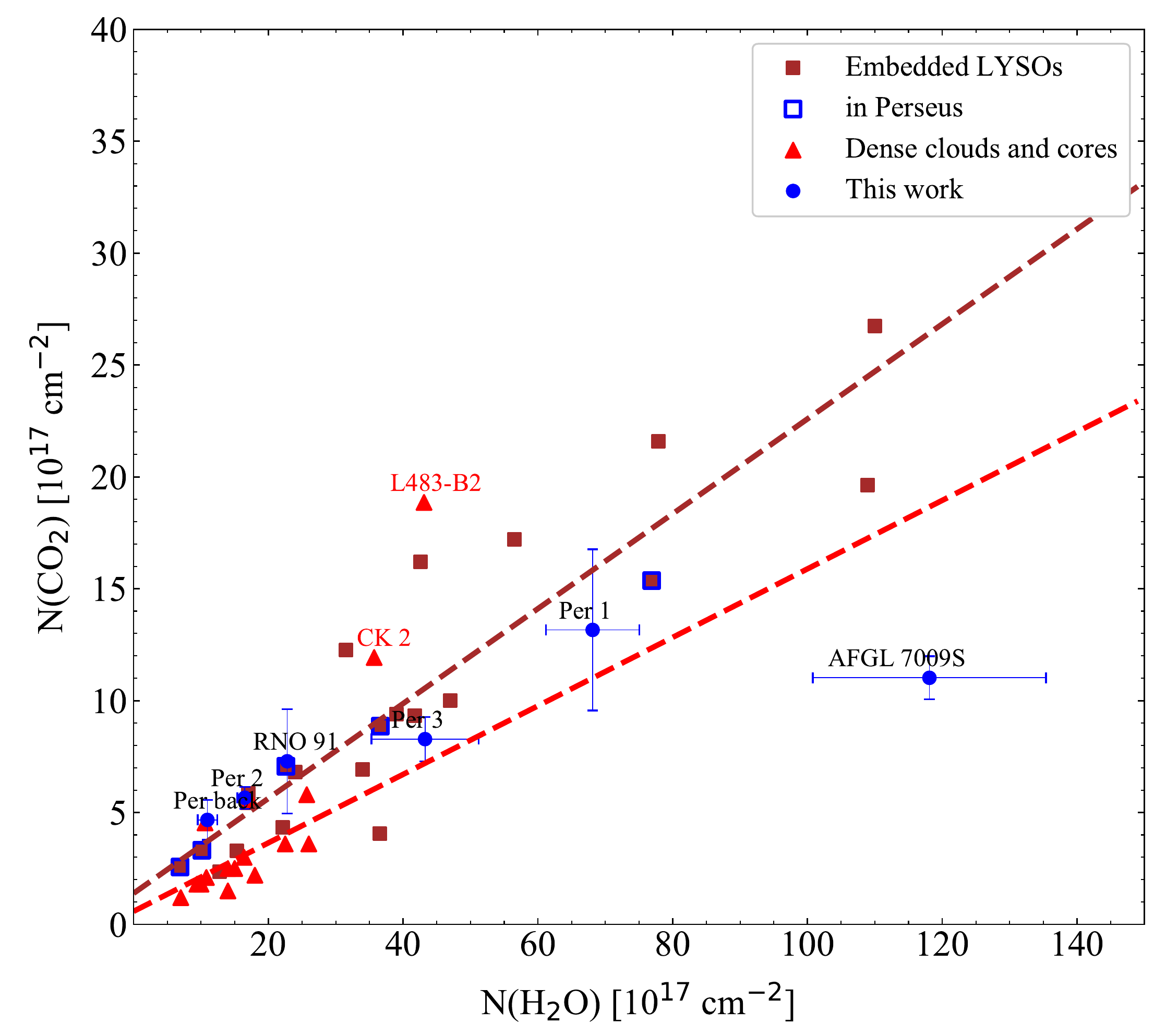}
\caption{The relation between column densities of \ch{H2O} ice and \ch{CO2} ice toward our targets (blue circles).
Brown squares are the embedded low-mass YSOs (LYSOs) from \citet{pon08b}, \citet{aik12}, and \citet{iop13}.
The brown squares framed with blue denote the embedded LYSOs in the Perseus molecular clouds.
The column densities of dense molecular clouds and cores (red triangles) are from \citet{whi07}, \citet{nob13}, and \citet{chu20}.
Each dashed line shows the best-fit results for the \ch{CO2} ice abundances of the embedded LYSOs (0.21), the molecular clouds and cores (0.15), respectively.}\label{fig7}
\end{center}
\end{figure}

\begin{figure}
\begin{center}
\includegraphics[scale=0.7]{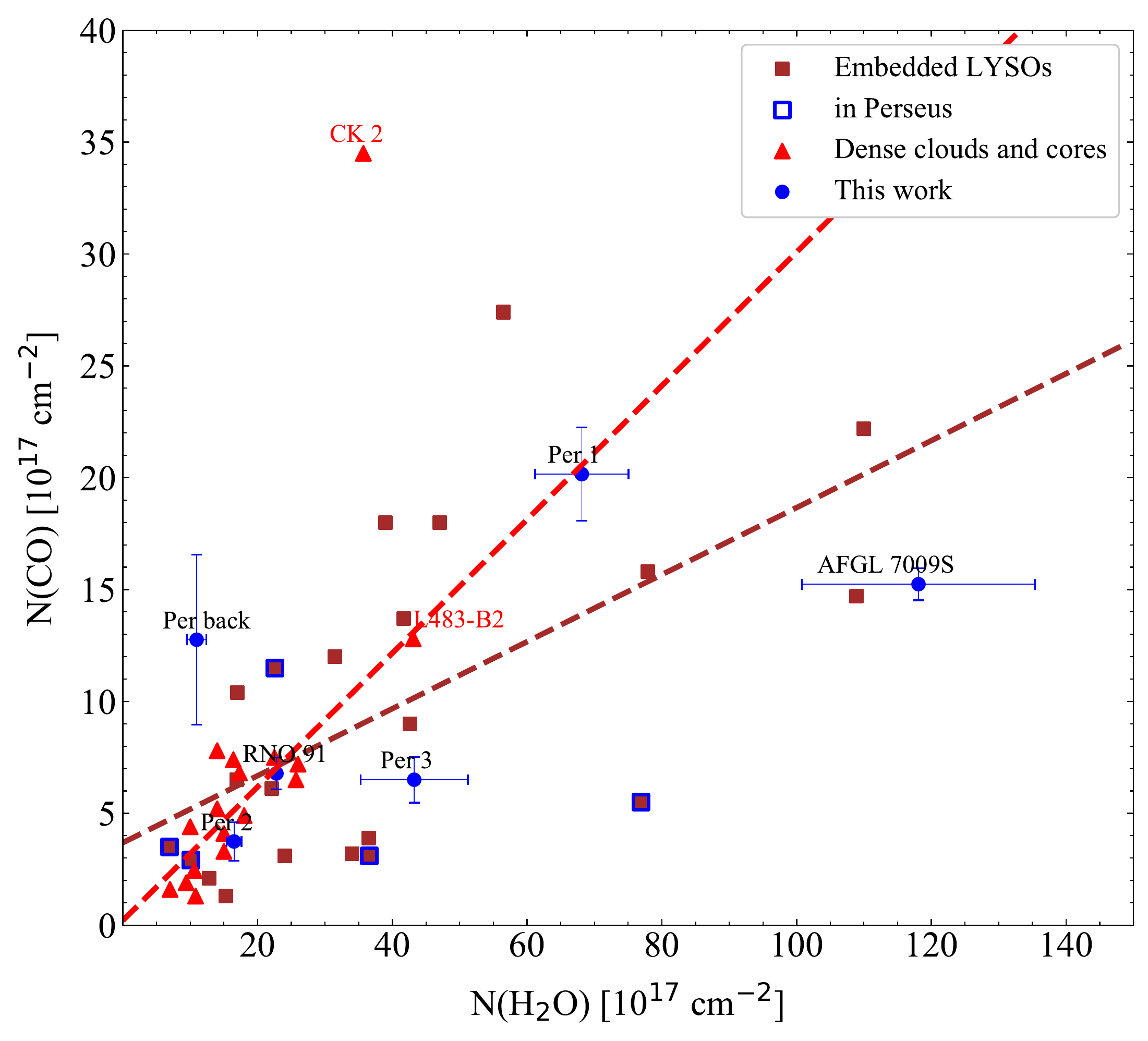}
\caption{The same plot as Figure~\ref{fig7}, but for the \ch{CO} ice.
Each dashed line shows the best-fit results for the \ch{CO} ice abundances of the embedded LYSOs (0.15), the molecular clouds and cores (0.30), respectively.}\label{fig8}
\end{center}
\end{figure}

To understand the chemical evolution of ice in the circumstellar envelope around low-luminosity protostars, we compared the ice abundances obtained from this study with those of dense molecular clouds and cores as well as other embedded low-mass young stellar objects (LYSOs).
In Figures ~\ref{fig7} and ~\ref{fig8}, we plot the \ch{CO2} and \ch{CO} ice abundances with respect to \ch{H2O} ice for our targets, LYSOs, and dense molecular clouds and cores together with their linear least square fitting results.
The ice abundances for the LYSOs were collected from the spectroscopic observational results with the VLT, IRTF, \textit{Spitzer IRS}, and \textit{AKARI IRC} \citep{pon03,pon08b,aik12,iop13}.
The selected LYSOs have evolutionary stages of \text{0} to \RomanNumeralCaps{1} \citep{iop13}.
The ice abundance data for the dense molecular cores were adopted from the \textit{AKARI} program toward field stars behind the quiescent cores \citep[A$_\mathrm{V}$ $>$ 6 mag;][]{nob13}.
The ice abundances of the dense molecular clouds ($A_\mathrm{V}$ $>$ 10 mag) were obtained from ground-based and \textit{Spitzer IRS} data toward field stars behind Serpens, Taurus, Lupus, and LDN483 regions \citep{whi07,boo11,boo13,iop13,chu20}.

Most of the LYSOs show higher values on the \ch{CO2} ice abundance than those of dense clouds and cores except for CK2 and L483-B2, which are denoted as red-colored text in Figure ~\ref{fig7}.
This trend can be explained as follows:
(1) most of the \ch{CO} gas in the pre-stellar core is frozen onto the cold grain surfaces, which are coated predominantly by \ch{H2O} ice. 
The reaction between \ch{CO} and \ch{OH} then leads to the formation of \ch{CO2} ice in a \ch{H2O}-rich ice mixture \citep{oba10,iop11}.
(2) the accretion rate of \ch{CO} gradually increases with the gas density, promoting the \ch{CO2} ice formation, and thus, increasing the amount of \ch{CO}-rich \ch{CO2} ice \citep{obe11,gar11} in a dense envelope. 
(3) in the protostellar stage, grain surfaces heated by protostellar radiation evaporate \ch{CO} molecules from the \ch{CO2}-containing ices, leaving the pure \ch{CO2} ice \citep{obe09}.
Indeed, \citet{pon08b} have detected the pure component of \ch{CO2} ice from the double-peaked absorption feature at 15.2 $\mu$m toward most of the observed LYSO targets.
As a result, the \ch{CO2} ice abundance increases as a protostar evolves.

For the background star behind the Perseus molecular cloud, the CO and \ch{CO2} ice abundances do not follow the trend of other molecular clouds and cores.
However, they have high values like the Serpens and LDN483 regions, which are traced by the spectra toward CK2 and L483-B2, respectively.
For the case of CK2, its high extincted line of sight ($A_\mathrm{V}$ $\sim$ 40 mag) and high \ch{CO2} ice abundance may indicate that the abundant gas-phase \ch{CO} molecules lead to sufficient \ch{CO2}-rich ice mixture formation, as well as the CO ice \citep{whi07}.
\citet{chu20} suggested that the environment for ice chemistry toward the dense region ($A_\mathrm{V}$ $\sim$ 40 mag) presented by L483-B2 may not be affected by the UV radiation or shocks from the nearby YSO.
Therefore, high CO and \ch{CO2} ice abundances for our background target may imply that the initial condition in the Perseus molecular cloud is comparable to those in the Serpens and LDN483 clouds.

For our low-luminosity targets, the \ch{CO2} ice abundance follows a trend similar to that of the protostars in the Perseus region (brown squares framed with blue).
This may indicate that their currently low luminosity is not an important parameter of the \ch{CO2} ice abundance.
As seen in Figure ~\ref{fig8}, however, their \ch{CO} ice abundances do not show a clear trend, unlike the \ch{CO2} ice abundance.
The \ch{CO} ice abundance in Perseus 1 is more similar to those in the molecular clouds and cores, where most \ch{CO} molecules are frozen on grain surfaces, rather than those of LYSOs, where \ch{CO} is in gas within the \ch{CO} sublimation radius.
However, Perseus 2 and 3 show a relatively low abundance of \ch{CO} ice, similar to that of AFGL 7009S, a massive protostar.
This result may suggest that they had a strong or recent burst event.
A strong burst can extend the \ch{CO} sublimation region to a large radius and retard the refreeze-out of the sublimated CO due to the low densities at large radii.
The deepest crystalline \ch{H2O} and \ch{XCN} ice features of Perseus 3 among the three low-luminosity targets support this interpretation.

\begin{figure}
\begin{center}
\includegraphics[scale=0.7]{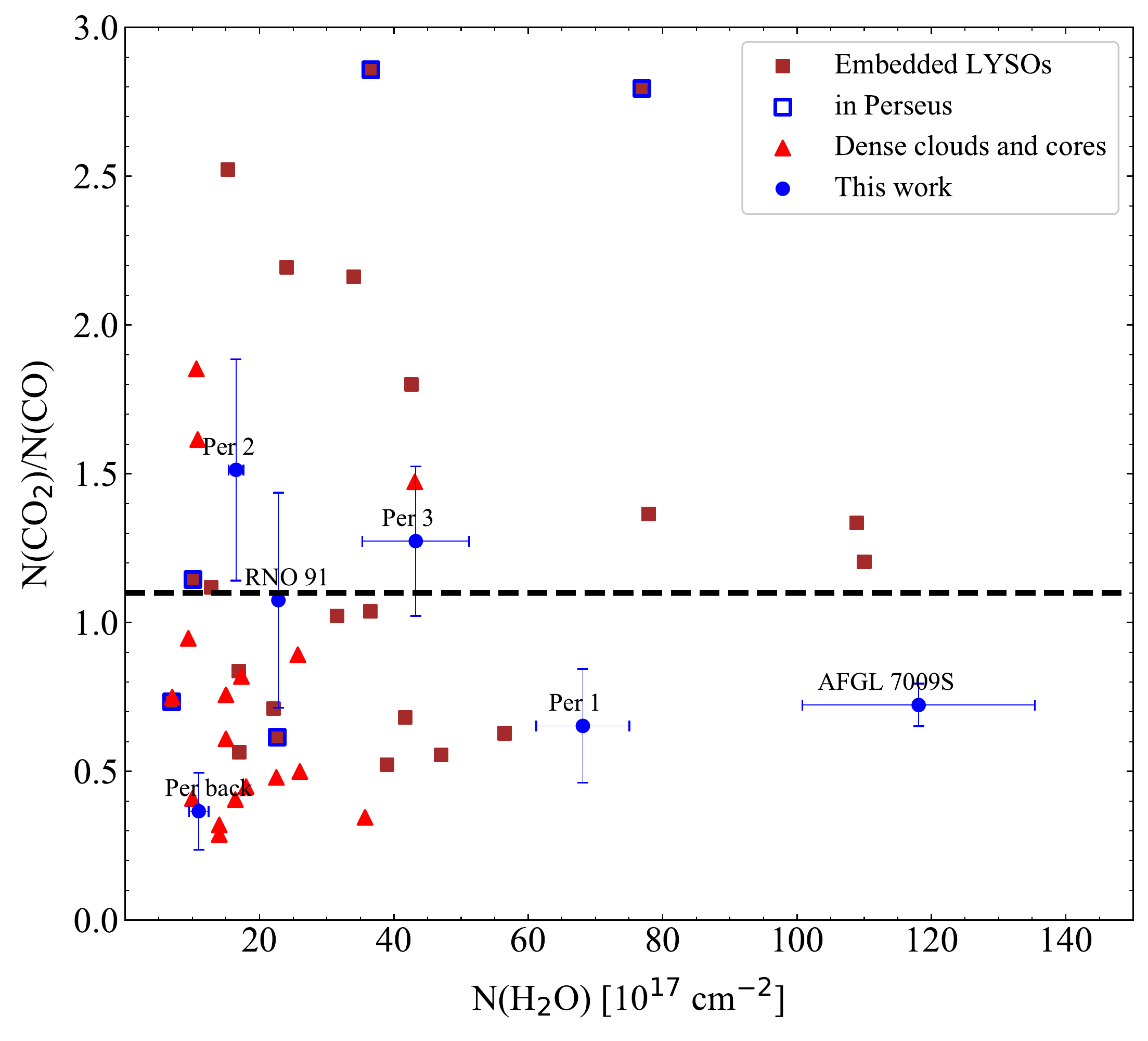}
\caption{The ice abundance ratio of \ch{CO2} to \ch{CO} ices versus \ch{H2O} ice column density.
Dashed horizontal line denotes the minimum ratio for the embedded LYSOs that have the double-peaked absorption feature of pure \ch{CO2} ice \citep[Figure 4 in][]{pon08b}.}\label{fig9}
\end{center}
\end{figure}

In the process of episodic accretion, the timescales of burst and quiescent phases are about 100 and 10,000 years, respectively \citep{vor13}. 
The sublimated \ch{CO2} during the burst events quickly returns to ices after the burst stops due to its higher binding energy\citep{vor13}.
On the other hand, the freeze-out timescale of CO is about 100,000 years at the density of $10^5$ cm$^{-3}$ \citep{lee04}. 
Therefore, some parts of CO sublimated during the previous burst events have not been fully frozen back onto grain surfaces. 
In Figure ~\ref{fig9}, most of the protostars with the double-peaked feature of pure \ch{CO2} ice (Figure 4 in \citet{pon08b}) have the ice abundance ratio of \ch{CO2} to \ch{CO} higher than 1.1.
It might imply that the protostellar heating by accretion episodes causes the decrease of \ch{CO} ice abundance \citep{obe11}.
Among our low-luminosity targets, Perseus 2 and 1 have the highest (1.5) and the lowest (0.7) \ch{CO2}/\ch{CO} ice ratios, respectively.
Considering the later evolutionary stage of Perseus 2 \citep[Class \RomanNumeralCaps{1}, Per-emb 38;][]{eno09} than Perseus 1 \citep[Class 0, Per-emb 25;][]{eno09}, Perseus 2 may have undergone more episodes of burst accretion, and more fraction of the sublimated CO ices may remain in the gas phase.

Recent ALMA survey of \ch{N2H+} (1$-$0) and \ch{HCO+} (3$-$2) toward protostars in the Perseus region have suggested timescales of the last burst event, using the emission peak radii of these two molecules \citep{hsi19}.
\ch{N2H+} and \ch{HCO+} peak radii reflect the CO and \ch{H2O} sublimation regions, where the gas-phase CO and \ch{H2O} destroy \ch{N2H+} and \ch{HCO+}, respectively \citep{vis15}.
Therefore, the estimated peak radius, which is larger than the expected value from the current luminosity, and the refreeze-out time of CO and \ch{H2O} provide the peak luminosity in the past burst and the timescale after the burst.
For Perseus 1 and 2, the \ch{HCO+} integrated intensity maps toward the central source enable measurement of the peak radius of the emission \citep[0.29$\arcsec$ and 0.05$\arcsec$, respectively; Figure 9 in][]{hsi19}.
Assuming an \ch{H2O} refreeze-out time of 1,000 years \citep{vis15}, \citet{hsi19} estimated the timescales after the last burst for Perseus 1 and 2 are shorter than 1,000 years and longer than 1,000 years, respectively.
However, for the case of Perseus 2, the detected \ch{HCO+} emission is too weak to estimate accurately its peak radius.
Furthermore, its high \ch{CO2}/\ch{CO} ice ratio and undetected \ch{N2H+} emission may suggest that a large amount of the sublimated \ch{CO} ices have not been frozen back onto grain surfaces.
Therefore, the last burst event for Perseus 2 should have occurred more recently than the case of Perseus 1.

\section{Summary}
We have carried out $AKARI$ IRC spectroscopic observations of three low-luminosity protostars in the Perseus molecular cloud to trace their thermal histories, printed in the ice compositions on dust grain surfaces.
We compared the ice abundances of these low-luminsity protostars with those of RNO 91, AFGL 7009S, and J03293654$+$3129465, which represent a low-mass protostar with a on-going burst accretion, a massive protostar, and a background star behind the Perseus molecular cloud, respectively.
All our targets show absorption features of \ch{H2O}, \ch{CO2}, and \ch{CO} ices around 3.05, 4.27, and 4.67 $\mu$m, respectively.
we estimated the ice column densities by fitting those ice features with the CDE-corrected laboratory data, and we also identified thermally produced ice components, such as XCN, during the past accretion bursts.

We compared the derived \ch{CO2} and \ch{CO} ice abundances of the low-luminosity targets with those of the embedded normal low-mass protostars and the dense molecular clouds and cores.
Our results are summarized as follows:
\paragraph{1.}The detection of the crystalline ice component of the \ch{H2O} absorption features toward the low-luminosity targets may support that they have experienced higher luminosity phases inducing crystallization of amorphous \ch{H2O} ices.
In the absorption band at 4.62 $\mu$m, the \ch{XCN} ice feature, which is a strong evidence of a past high-luminosity stage in currently low-luminosity protostars, was clearly detected in at least two targets (Perseus 1 and Perseus 3).
\paragraph{2.}The \ch{CO2} ice abundances of our low-luminosity targets show a trend similar to those of the embedded low-mass protostars, indicating a common formation process of \ch{CO2} ice in the protostellar evolution and insensitive \ch{CO2} ice abundance to the accretion history.
\paragraph{3.}The CO ice abundances of our low-luminosity targets show scattered distribution, unlike the \ch{CO2} ice abundance.
This result implies that different thermal conditions by past accretion bursts and the timescale after the last burst have contributed to the refreeze-out of the sublimated CO.
\paragraph{4.}Therefore, the \ch{CO2}/CO ice ratio can be used to estimate the timescale after the last accretion burst event as the combination of the \ch{HCO+} and \ch{N2H+} emission peaks does.
\paragraph{5.}For the case of Perseus 3, the past burst event is suspected to be much stronger than other low-luminosity targets, resulting in the deepest crystalline \ch{H2O} and XCN ice features.
\paragraph{6.}The CO and \ch{CO2} ice abundances in the Perseus molecular cloud are relatively higher than those obtained from other dense molecular clouds and cores, probably because the abundant CO molecules promote those ice formation sufficiently in the Perseus molecular cloud.

This work was supported by Basic Science Research Program through the National Research Foundation of Korea(NRF) funded by the Ministry of Education(grant number 2018R1A6A3A01013296).
This work was supported by the National Research Foundation of Korea (NRF) grant funded by the Korea government (MSIT) (grant number 2021R1A2C1011718).
This work is based on observations with \textit{AKARI}, a JAXA project with the participation of ESA.
This publication makes use of data products from the Two Micron All Sky Survey.
The Two Micron All Sky Survey is a joint project of the University of Massachusetts and the Infrared Processing and Analysis Center/California Institute of Technology, funded by the National Aeronautics and Space Administration and the National Science Foundation.
This research has made use of the NASA/ IPAC Infrared Science Archive, which is operated by the Jet Propulsion Laboratory, California Institute of Technology, under contract with the National Aeronautics and Space Administration.

\clearpage

\begin{deluxetable}{ccccccccccccccccccc}
\setlength{\tabcolsep}{0.02in}
\tablewidth{0pt}
\tablecaption{Photometric information \label{tab-1}}

\tablehead{
\colhead{Source name} & \multicolumn{3}{c}{Position} & \colhead{} & \multicolumn{11}{c}{Fluxes (mJy)} & \colhead{} & \colhead{L$_{bol}$\tablenotemark{a}} & \colhead{T$_{bol}$\tablenotemark{a}} \\
\colhead{} & \colhead{} & \colhead{} & \colhead{} & \colhead{} &
\multicolumn{3}{c}{2MASS} & \colhead{} & \multicolumn{4}{c}{IRAC Band} & \colhead{} & \multicolumn{2}{c}{MIPS Band} &
\colhead{} & \colhead{} & \colhead{} \\
\cline{2-4} \cline{6-8} \cline{10-13} \cline{15-16}
\colhead{} & \colhead{${{\alpha_{J2000.0}}}$} & \colhead{} & \colhead{${\delta_{J2000.0}}$} &
\colhead{} & \colhead{$J$} & \colhead{$H$} & \colhead{$K$} & \colhead{} &
\colhead{1} & \colhead{2} & \colhead{3} & \colhead{4} & \colhead{} &
\colhead{1}&\colhead{2}& \colhead{} &
\colhead{($L_{\astrosun}$)}&\colhead{($K$)}
}

\startdata
Perseus 1 & 3 26 37.46 && 30 15 28.1 && 0.18 & 0.50 & 1.19 && 3.62 & 9.27 & 11.1 & 12.0 && 396 & 4300 && 1.10 & 64 \\
          &            &&            &&      &      &      &&      &      &      &      &&     &     && (0.69\tablenotemark{b}) & \\
Perseus 2 & 3 32 29.18 && 31 02 40.9 && 0.71 & 2.69 & 5.81 && 11.1 & 17.2 & 23.3 & 32.8 && 181 & 1130 && 0.57 & 114 \\
          &            &&            &&      &      &      &&      &      &      &      &&     &     && (0.20\tablenotemark{b}) & \\
Perseus 3 & 3 43 51.02 && 32 03 07.9 && $\cdots$ & $\cdots$ & $\cdots$ && 6.81 & 3.99 & 12.8 & 2.15 && 45.4 & 1960 && 0.32 & 63 \\
          &            &&            &&      &      &      &&      &      &      &      &&     &     && (0.33\tablenotemark{b}) & \\
J03293654+3129465 & 3 29 36.54 && 31 29 45.5 && 13.4 & 76.6 & 130 && 102 & 69.8 & 53.6 & 29.9 && 3.75 & $\cdots$ && $\cdots$ & $\cdots$ \\
RNO 91 & 16 34 29.32 && -15 47 01.4 && 63.8 & 143 & 274 && 737 & 1110 & 1440 & 1660 && 5000 & 20300 && 2.6\tablenotemark{c} & 340\tablenotemark{c} \\
AFGL 7009S & 18 34 20.91 && -5 59 42.2 && 0.13 & 0.50 & 9.00 && $\cdots$ & $\cdots$ & $\cdots$ & $\cdots$ && $\cdots$ & $\cdots$ && 2.9$\times$10$^{4}$\tablenotemark{d} & $\cdots$\\
\enddata
%
%

\tablenotetext{a}{$L_{bol}$ and T$_{bol}$ are referred from \citet{dun08}, excepting for RNO91 and AFGL 7009S.}
\tablenotetext{b}{Internal luminosity ($L_{int}$) of a central protostar. $L_{int}$ is derived from 70 $\mu$m flux \citep{dun08}}
\tablenotetext{c}{\citet{lee15b}}
\tablenotetext{d}{\citet{zap09}}
\end{deluxetable}
\clearpage

\begin{deluxetable}{ccccccc}
\tablecaption{Best fitting laboratory data\tablenotemark{a} \label{tab-2}}
\tablewidth{0pt}
\setlength{\tabcolsep}{0.05in}
\tablehead{\colhead{Source name} & \colhead{} & \colhead{Polar component} & \colhead{} & \colhead{Apolar component} & \colhead{} & \colhead{Pure component\tablenotemark{b}}}

\startdata
Perseus 1 && \ch{H2O}:\ch{NH3} $=$ 100:20 && \ch{CO}:\ch{CO2} $=$ 100:70 && \ch{H2O}, \ch{CO2} \\
          && \ch{H2O}:\ch{CO2} $=$ 100:14 &&  &&  \\
          && \ch{H2O}:\ch{CO} $=$ 100:20 &&  &&  \\
Perseus 2 && \ch{H2O}:\ch{NH3} $=$ 100:20 && \ch{CO}:\ch{CO2} $=$ 100:70 && \ch{H2O}, \ch{CO2} \\
          && \ch{H2O}:\ch{CO} $=$ 100:20 && \ch{H2O}:\ch{CO2} $=$ 1:6 && \\
Perseus 3 && \ch{H2O}:\ch{CO} $=$ 100:10 && \ch{CO}:\ch{CO2} $=$ 100:70 && \ch{H2O}, \ch{CO2} \\
          &&                             && \ch{H2O}:\ch{CO2} $=$ 1:6 &&  \\
J03293654+3129465 && \ch{H2O}:\ch{NH3} $=$ 100:20 && \ch{H2O}:\ch{CO2} $=$ 1:6 && \ch{H2O} \\
                  && \ch{H2O}:\ch{CO} $=$ 100:20 && \ch{CO}:\ch{CO2} $=$ 1:1 at 15 K && \\
RNO 91 && \ch{H2O}:\ch{NH3} $=$ 100:20 && \ch{H2O}:\ch{CO2} $=$ 1:6 && \ch{H2O}, \ch{CO2} \\
       && \ch{H2O}:\ch{CO} $=$ 100:20 && \ch{CO}:\ch{CO2} $=$ 100:70 && \\
AFGL 7009S && \ch{H2O}:\ch{NH3} $=$ 100:20 && \ch{H2O}:\ch{CO2} $=$ 1:6 && \ch{H2O} \\
           && \ch{H2O}:\ch{CO2}:\ch{CO} $=$ 100:20:3 at 20 K && \ch{H2O}:\ch{CO}:\ch{O2} $=$ 1:20:60 at 30 K && \\
           &&                                                && \ch{H2O}:\ch{CH3OH}:\ch{CO2} $=$ 1:1:1 at 112 K && \\
\enddata

\tablenotetext{a}{All ice mixtures are the results of experiments on 10 K except the cases of AFGL 7009S and J03293654+3129465.}
\tablenotetext{b}{Pure \ch{H2O} ices are divided into amorphous (10 K) and crystralline (160 K) components except the cases of AFGL 7009S (amorphous component at 50 K only) and J03293654+3129465 (amorphous component at 10 K only). All pure \ch{CO2} profiles are the results of experiments on 20 K.}
\end{deluxetable}
\clearpage

\begin{deluxetable}{ccccccccccc}
\tablecaption{Ice column densities \label{tab-3}}
\tablewidth{0pt}
\setlength{\tabcolsep}{0.05in}
\tablehead{\colhead{Source name} & \colhead{} & \colhead{N(\ch{H2O})} & \colhead{} & \colhead{N(\ch{CO2})} & \colhead{} & \colhead{N(\ch{CO})} & \colhead{} & \colhead{N(\ch{XCN})\tablenotemark{a}} & \colhead{} & \colhead{N(\ch{OCS})\tablenotemark{b}} \\
\colhead{} & \colhead{} & \colhead{$\times$10$^{17}$ (cm$^{-2}$)} & \colhead{} & \colhead{$\times$10$^{17}$ (cm$^{-2}$)} & \colhead{} & \colhead{$\times$10$^{17}$ (cm$^{-2}$)} & \colhead{} & \colhead{$\times$10$^{17}$ (cm$^{-2}$)} & \colhead{} & \colhead{$\times$10$^{17}$ (cm$^{-2}$)}}

\startdata
Perseus 1 && 68.12 $\pm$ 6.92 && 13.16 $\pm$ 3.60 && 20.16 $\pm$ 2.09 && 0.28 $\pm$ 0.31 && 0.43 $\pm$ 0.06\\
          &&                  && 28.62 $\pm$ 1.21 \tablenotemark{c} &&                  &&                 &&  \\ 
Perseus 2 && 16.50 $\pm$ 1.11 && 5.66 $\pm$ 0.49 && 3.74 $\pm$ 0.86 && 0.05 $\pm$ 0.18 && 0.12 $\pm$ 0.05\\
          &&                  && 5.04 $\pm$ 0.13 \tablenotemark{c} &&                  &&                 &&  \\ 
Perseus 3 && 43.25 $\pm$ 7.96 && 8.28 $\pm$ 0.99 && 6.50 $\pm$ 1.02 && 0.56 $\pm$ 0.22 && 0.24 $\pm$ 0.08\\
J03293654+3129465 && 10.65 $\pm$ 1.45 && 4.67 $\pm$ 0.89 && 12.76 $\pm$ 3.80 && $\cdots$ && 0.26 $\pm$ 0.05\\
RNO 91 && 22.80 $\pm$ 0.25 && 7.29 $\pm$ 2.33 && 6.78 $\pm$ 0.71 && 0.33 $\pm$ 0.46 && 0.11 $\pm$ 0.04\\
       && 39.00 $\pm$ 5.00 \tablenotemark{c} && 11.66 $\pm$ 0.16 \tablenotemark{c} && 8.00 $\pm$ 0.30 \tablenotemark{c} &&                && \\
AFGL 7009S && 118.1 $\pm$ 17.30 && 11.02 $\pm$ 0.96 && 15.24 $\pm$ 0.72 && 1.21 $\pm$ 0.44 && 0.34 $\pm$ 0.04\\
           && 110 \tablenotemark{d}               && 25 \tablenotemark{d}               && 18 \tablenotemark{d}               && 4.2 \tablenotemark{d}             && 0.11 \tablenotemark{d}\\
\enddata

\tablenotetext{a}{\ch{XCN} column density is based on the assumption that \ch{XCN} is \ch{OCN-}. The band strength of \ch{OCN-} is 5$\times$10$^{-17}$ cm molecule$^{-1}$ \citep{sch97}.}
\tablenotetext{b}{The band strength of \ch{OCS} is 1.5$\times$10$^{-16}$ cm molecule$^{-1}$ \citep{pal97}.}
\tablenotetext{c}{Previous ground-based and \textit{Spitzer} IRS results \citep{pon08b,kim12,iop13} for our targets.}
\tablenotetext{d}{Previous \textit{ISO} results \citep{gib04} for AFGL 7009S.}
\end{deluxetable}
\clearpage

\begin{deluxetable}{ccccc}
\tablecaption{Ice abundances\label{tab-4}}
\tablewidth{0pt}
\setlength{\tabcolsep}{0.05in}
\tablehead{\colhead{Source name} & \colhead{} & \multicolumn{3}{c}{Ice abundances} \\
\cline{3-5}
\colhead{} & \colhead{} & \colhead{X(\ch{CO2}$_{ice}$)\tablenotemark{a}} & \colhead{} & \colhead{X(\ch{CO}$_{ice}$)\tablenotemark{b}}}

\startdata
Perseus 1 && 0.19 $\pm$ 0.06 $-$ 0.42 $\pm$ 0.05 \tablenotemark{c} && 0.30 $\pm$ 0.04 \\
Perseus 2 && 0.31 $\pm$ 0.02 \tablenotemark{c} $-$ 0.34 $\pm$ 0.04 && 0.23 $\pm$ 0.05 \\
Perseus 3 && 0.19 $\pm$ 0.04 && 0.15 $\pm$ 0.04 \\
J03293654+3129465 && 0.43 $\pm$ 0.10 && 1.17 $\pm$ 0.38 \\
RNO 91 && 0.30 $\pm$ 0.04 \tablenotemark{d} $-$ 0.32 $\pm$ 0.10 && 0.21 $\pm$ 0.03 \tablenotemark{d} $-$ 0.30 $\pm$ 0.03 \\
AFGL 7009S && 0.09 $\pm$ 0.02 $-$ 0.23 \tablenotemark{e} && 0.13 $\pm$ 0.02 $-$ 0.16 \tablenotemark{e} \\
Embedded LYSOs && 0.21 $\pm$ 0.02 && 0.15 $\pm$ 0.04 \\
Dense clouds and cores && 0.15 $\pm$ 0.05 && 0.30 $\pm$ 0.08\\
\enddata

\tablenotetext{a}{X(\ch{CO2}$_{ice}$) $=$ N(\ch{CO2})/N(\ch{H2O})}
\tablenotetext{b}{X(\ch{CO}$_{ice}$) $=$ N(\ch{CO})/N(\ch{H2O})}
\tablenotetext{c}{Previous \textit{Spitzer} IRS results \citep{kim12} for Perseus 1 and Perseus 2. The \ch{H2O} ice column densities from this work were adopted to calculate X(\ch{CO2}$_{ice}$).}
\tablenotetext{d}{Previous ground-based and \textit{Spitzer} IRS results \citep{pon08b,iop13} for RNO 91.}
\tablenotetext{e}{Previous \textit{ISO} results \citep{gib04} for AFGL 7009S.}

\end{deluxetable}
\clearpage

\end{document}